\UseRawInputEncoding
\documentclass[conference]{IEEEtran}
\IEEEoverridecommandlockouts
\usepackage{cite}
\usepackage{amsmath,amssymb,amsfonts}
\usepackage{algorithmic}
\usepackage{graphicx}
\usepackage{textcomp}
\usepackage{xcolor}
\usepackage[ruled,vlined]{algorithm2e}
\usepackage{float}
\def\BibTeX{{\rm B\kern-.05em{\sc i\kern-.025em b}\kern-.08em
    T\kern-.1667em\lower.7ex\hbox{E}\kern-.125emX}}

\usepackage{booktabs} 

\usepackage{color, colortbl}
\definecolor{Gray}{gray}{0.9}
\definecolor{LightCyan}{rgb}{0.88,1,1}
\usepackage[first=0,last=9]{lcg}

\usepackage{xcolor}

\begin{document}

\title{Deep Learning Based Antenna Selection Technique for RIS-Empowered RQSM System \\
\thanks{This work was supported by The Scientific and Technological Research Council of Turkey (TUBITAK) through the 1515 Frontier Research and Development Laboratories Support Program under Project 5229901 - 6GEN. Lab: 6G and Artificial Intelligence Laboratory. Also, this work was supported by TUBITAK 1001 (Grant Number: 123E513).}
}

\author{\IEEEauthorblockN{Burak Ahmet Ozden \IEEEauthorrefmark{1}\IEEEauthorrefmark{2},
Fatih Cogen \IEEEauthorrefmark{3},
Erdogan Aydin\IEEEauthorrefmark{1}}

\IEEEauthorblockA{\IEEEauthorrefmark{1}Department of Electrical and Electronics Engineering, Istanbul Medeniyet University, Istanbul, Turkey.}
\IEEEauthorblockA{\IEEEauthorrefmark{2}Department of Computer Engineering, Yildiz Technical University, Istanbul, Turkey.}
\IEEEauthorblockA{\IEEEauthorrefmark{3}Turkcell 6GEN. Lab
Turkcell Iletisim Hizmetleri Inc., Istanbul, Turkey.}
\vspace{-0.3cm}
\\  Email: bozden@yildiz.edu.tr, fatih.cogen@turkcell.com.tr,  erdogan.aydin@medeniyet.edu.tr.
}

\maketitle

\begin{abstract}
Reconfigurable intelligent surface (RIS) technology has attracted considerable interest due to its ability to control wireless propagation with minimal power usage. Receive quadrature spatial modulation (RQSM) scheme transmits data bits in both in-phase ($I$) and quadrature ($Q$) channels, doubling the number of active receive antenna indices and improving spectral efficiency compared to the traditional receive spatial modulation (RSM) technique. Also, capacity-optimized antenna selection (COAS) improves error performance by selecting antennas with the best channel conditions. This paper proposes a new deep neural network (DNN)-based antenna selection method, supported by the COAS technique, to improve the error performance of the RIS-RQSM system. Monte Carlo simulations of the proposed DNN-COAS-RIS-RQSM system using the quadrature amplitude modulation (QAM) technique for Rayleigh fading channels are performed and compared with the COAS-RIS-RQSM system. Also, a comparative analysis of the computational complexities of the DNN and COAS techniques is conducted to evaluate the trade-offs between error performance and complexity.
\end{abstract} 


\section{Introduction}

Future wireless networks are expected to support multiple connected devices simultaneously, deliver unprecedented data rates, and meet stringent requirements for low latency and energy efficiency. These demands are driven by emerging use cases such as immersive augmented/virtual reality experiences requiring ultra-low latency and high data rates, high-precision industrial automation with real-time control loops, massive Internet of Things (IoT) ecosystems leveraging data-driven intelligence, and expanding smart city infrastructures requiring robust, scalable connectivity solutions. Traditional communication paradigms often struggle to meet these performance metrics due to physical and operational constraints such as limited spectrum, multipath fading, and high hardware complexity \cite{Qadir_Towards_6g, 6Gs}. Therefore, researchers are actively pursuing innovations to reshape the propagation environment, optimize resource allocation, and exploit additional information dimensions with minimal energy consumption. In this context, reconfigurable intelligent surfaces (RISs), advanced index modulation (IM) schemes, and deep neural network (DNN)-based antenna selection methods show strong potential.

Antenna selection techniques improve the performance of multi-antenna wireless communication systems by selecting the subset of antennas with the best channel conditions from the available antennas. Capacity-optimized antenna selection (COAS) is a technique in wireless communication systems that selects a subset of antennas to maximize channel capacity under given channel conditions \cite{Siu2024-fatih, ASburak}. However, the improvement in error performance provided by the COAS technique is insufficient to meet the requirements of modern wireless communication systems. Recent studies have increasingly employed DNNs to address the antenna selection problem \cite{dnn_sm1,dnn_sm2}. They indicate that DNN-based methods can learn from channel state distributions and efficiently select effective antenna subsets. Also, the effectiveness of integrating COAS with DNNs in spatial modulation (SM) system has been extensively demonstrated by Cogen \textit{et al.} at \cite{fatih-DNN-ELECO}. The proposed DNN-based system achieves high energy efficiency and significant performance gains compared to classical SM schemes, as verified by simulation results.

\begin{figure*}[t]
\centering{\includegraphics[width=1\textwidth]{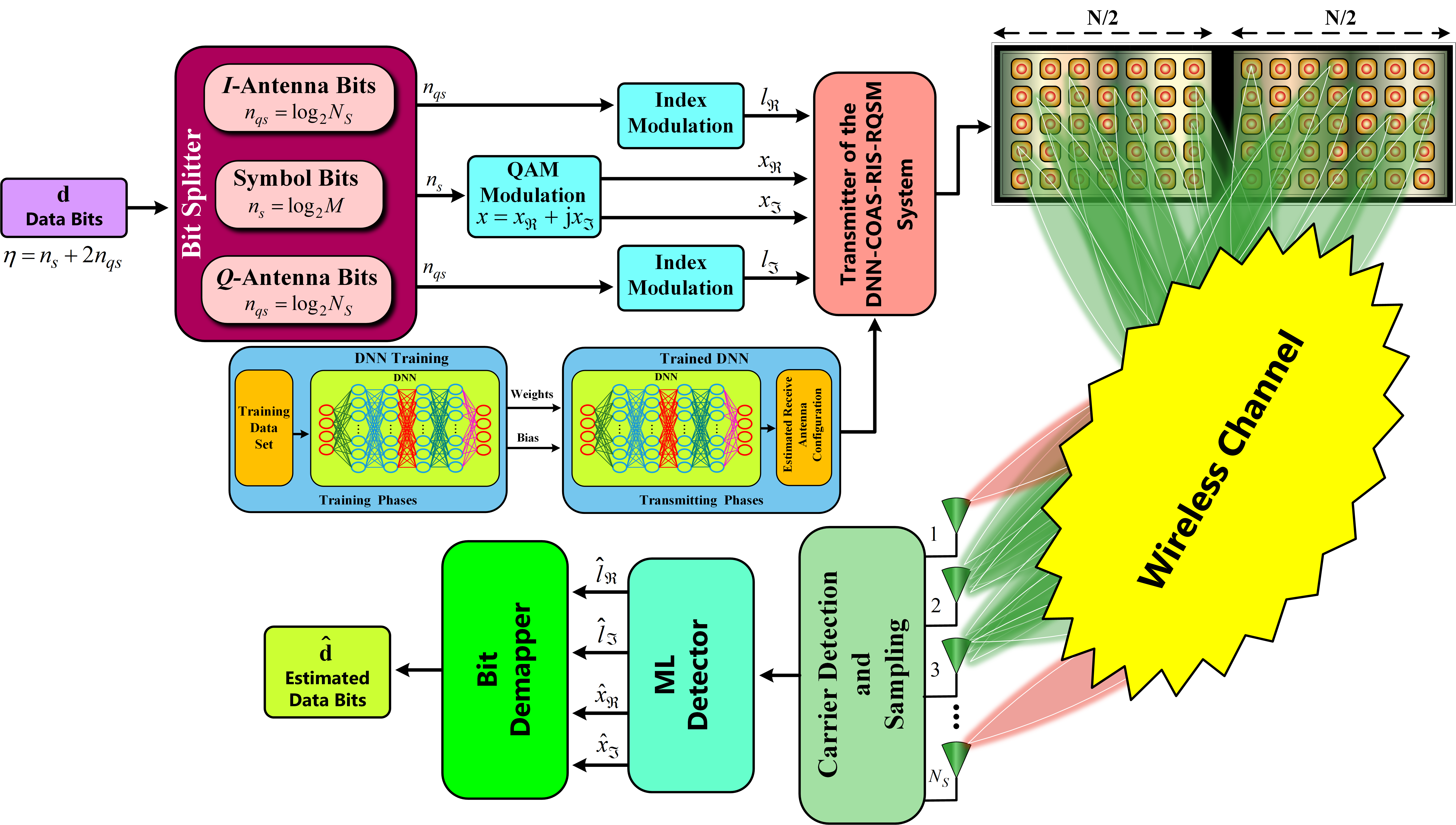}}	
	\caption{System model of the proposed DNN-COAS-RIS-RQSM system.}
	\label{system_model} 
\end{figure*}

RISs have attracted attention as a promising technology to improve radio resource management. By intelligently adjusting the phase shifts of passive reflective elements, RIS can dynamically change the wireless propagation environment to enhance signal quality and manage interference. Unlike traditional relay-based systems, RISs require minimal power, involve less hardware complexity, and can be easily integrated into existing infrastructures such as walls, facades, and ceilings \cite{RIS1, Fatih-ICCSPA, RIS_BURAK, wcnc-dubai, CIM-RIS-Fatih}. Recent studies have also used machine learning methods to automatically control RIS reflections in real-time, opening the door to fully adaptive networks that can flexibly adjust radio environments to meet demanding quality of service (QoS) requirements in different situations \cite{RIS-farklı-survey, RISmac}.

SM is a popular IM technique that improves spectral efficiency and error performance compared to conventional communication techniques by embedding additional information in the antenna indices. Unlike V-BLAST, which can activate multiple antennas simultaneously, SM activates only one antenna in each transmission interval. This eliminates inter-channel interference and the need for synchronization between antennas and reduces complexity \cite{smsurvey, mesleh-ilk-SM, CIMSMBM}. However, further increases in data rates of the SM technique are possible. The quadrature SM (QSM) technique extends the concept of SM by splitting the transmitted signal into two spatial dimensions—namely in-phase and quadrature channels—and enabling the simultaneous activation of two antennas on orthogonal sub-channels \cite{qsm1}.
This approach doubles the information conveyed by the antenna indices while retaining the advantages of the SM technique. Consequently, the QSM technique significantly enhances spectral efficiency and error performance compared to the conventional SM technique.

\subsection{Contribution} 
This paper introduces a new high-performance, energy-efficient DNN-based wireless communication architecture, termed DNN-COAS-RIS-RQSM, which integrates RQSM, COAS, and RIS within a DNN-based framework. Keeping all other system parameters constant, the number of receive antennas and reflecting elements is varied to evaluate their effect on bit error rate (BER) performance. Simulation results show that the proposed DNN-COAS-RIS-RQSM system achieves better error performance than the conventional COAS-RIS-RQSM technique. In addition to evaluating performance, the computational complexity of the DNN is compared with that of the traditional COAS technique to highlight the trade-off between complexity and detection accuracy.

\subsection{Organization} 
The remainder of this paper is organized as follows. Section II presents the system model of the proposed DNN-COAS-RIS-RQSM scheme. Section III presents complexity analyses of DNN and COAS techniques. Section IV offers simulation results. Finally, Section V concludes the paper.

\subsection{Notation} 
Vectors and matrices are denoted by bold lowercase (e.g., \(\mathbf{x}\)) and bold uppercase letters (e.g., \(\mathbf{X}\)), respectively. The transpose of a matrix or vector \(\mathbf{X}\) is represented by \(\mathbf{X}^T\). The notation \(\lvert x \rvert\) indicates the absolute value of a scalar \(x\). The Euclidean norm of a vector \(\mathbf{x}\) is denoted by \(\lVert \mathbf{x} \rVert\), while the Frobenius norm of a matrix \(\mathbf{X}\) is denoted by \(\lVert \mathbf{X} \rVert_F\). The real and imaginary parts of a complex number \(x\) are denoted by $x_\Re$ and $x_\Im$, respectively.

\section{System Model}
Fig. \ref{system_model} presents the system model of the proposed DNN-COAS-RIS-RQSM scheme, which is designed by combining RIS technology, COAS-assisted DNN-based antenna selection, and the QSM technique, which is high data rate index modulation, within a SIMO structure. Since the transmitter of the proposed system is positioned very close to the RIS, it is assumed that there is no channel between the transmitter and the RIS \cite{RIS1}. The proposed system model consists of $N_S$ receive antennas and $N$ reflecting surfaces. Half of these reflecting elements belong to the first RIS unit, while the remaining half constitutes the second RIS unit.  The proposed system transmits data bits over the receive antenna indices and QAM symbols. Also, the RIS directs the signals arriving at its surfaces to the receive antennas, which are selected with a phase rotation that maximizes the SNR of the transmitted signals. The spectral efficiency of the proposed system is given as follows:
\begin{eqnarray}\label{spectral}
\eta= \log_2(M) + 2\log_2(N_S),
\end{eqnarray}
where $M$ is the modulation order of the QAM. 

In the system model of the proposed scheme, in a transmission period, $\eta$ bits in the vector $\textbf{d}$ are transmitted to the receiver via RIS. Of these $\eta$ bits, the $n_s=\log_2(M)$ bit selects the transmitted $M$-QAM symbol $x=x_\Re+jx_\Im$. Of the remaining $2n_{qs}=2\log_2(N_S)$ bits, half \big($n_{qs}=\log_2(N_S)$ bits\big) selects the $l_\Re$th receive antenna corresponding to the $I$ component, and the other half $n_{qs}$ bits selects the $l_\Im$th  receive antenna corresponding to the $Q$ component. Therefore, the receive antennas $l_\Re$, $l_\Im$, and symbols $x_\Re$, $x_\Im$ are selected according to the data bits in the transmission period. 

The channel matrix $\textbf{H} \in \mathbb{C}^{N\times N_R}$ of the traditional RIS-RQSM scheme is expressed as follows:
\begin{eqnarray}\label{channel_matrix1}
\textbf{H} &=&\begin{bmatrix}
h_{1,1} & h_{1,2} &  \cdots  & h_{1,N_R}  \\ 
h_{2,1} &h_{2,2} & \cdots  & h_{2,N_R}  \\ 
\vdots & \vdots & \ddots& \vdots\\ 
h_{N,1} & h_{N,2} & \cdots  & h_{N,N_R}
\end{bmatrix}_{N \times N_R}\!\!\!\!\!\!\!\!\!\!\!\!.
\end{eqnarray}

In the traditional COAS-RIS-RQSM system,  a subset of $N_S$ antennas is selected from $N_R$ receive antennas using the COAS technique. The norms of the $N_R$ channel coefficients, ranked in descending order, are expressed as follows:
\begin{eqnarray}\label{coaseq1}
         \big|\big|\mathbf{h}_1\big|\big|^2 \geq \ldots \geq \big|\big|\mathbf{h}_{N_S}\big|\big|^2 \geq \big|\big|\mathbf{h}_{N_{S+1}}\big|\big|^2 \geq \ldots \geq \big|\big|\mathbf{h}_{N_R}\big|\big|^2.
\end{eqnarray}
In (\ref{coaseq1}), the selected channel matrix \( \mathbf{H}_{\mathcal{S}} = \left[\mathbf{h}_1, \mathbf{h}_2, \ldots,\mathbf{h}_{N_{S}}  \right] \in \mathbb{C}^{N\times N_S} \) is constructed by selecting the \( N_S \) channel vectors with the largest norms from the set of channel coefficient vectors, which are initially sorted in ascending order based on their norms. With the first $N/2$ columns of the selected channel matrix $\textbf{H}_S \in \mathbb{C}^{N\times N_S}$, the channel matrix $\textbf{H}_S^\Re \in \mathbb{C}^{N/2\times N_S}$ corresponding to the in-phase ($I$) component is formed and with the other $N/2$ columns, the channel matrix $\textbf{H}_S^\Im \in \mathbb{C}^{N/2\times N_S}$ corresponding to the quadrature ($Q$) component is formed, i.e. $\textbf{H}_S=[\textbf{H}_S^\Re \ \textbf{H}_S^\Im]$. The $\textbf{H}_S^\Re$ and $\textbf{H}_S^\Im$ can be expressed as follows:
\begin{eqnarray}\label{channel_matrix1_2}
\textbf{H}_S^\Re &=&\begin{bmatrix}
h_{1,1} & h_{1,2} &  \cdots  & h_{1,N_S}  \\ 
h_{2,1} &h_{2,2} & \cdots  & h_{2,N_S}  \\ 
\vdots & \vdots & \ddots& \vdots\\ 
h_{N/2,1} & h_{N/2,2} & \cdots  & h_{N/2,N_S}
\end{bmatrix}_{\frac{N}{2} \times N_S}\!\!\!\!\!\!\!\!\!\!\!\!,
\end{eqnarray}
\begin{eqnarray}\label{channel_matrix1_3}
\textbf{H}_S^\Im =\begin{bmatrix}
\!\!\!h_{1,1} & \!\!\! h_{1,2} &  \!\!\!\!\!\!\cdots  & \!\!\!\!\!\! h_{1,N_S}  \\ 
\!\!\!h_{2,1} &\!\!\!h_{2,2} & \!\!\!\!\!\! \cdots  & \!\!\!\!\!\! h_{2,N_S}  \\ 
\!\!\!\!\!\!\vdots & \!\!\! \vdots & \!\!\! \!\!\!\ddots& \!\!\! \vdots\\ 
h_{(N/2+1),1} & h_{(N/2+1),2} & \!\!\! \cdots  & \!\!\! h_{(N/2+1),N_S}
\end{bmatrix}_{\frac{N}{2} \times N_S}\!\!\!\!\!\!\!\!\!\!\!\!\!\!\!\!\!\!\!\!\!\!.
\end{eqnarray}
The selected channel matrices in (\ref{channel_matrix1_2}) and (\ref{channel_matrix1_3}) are used in (\ref{input_v}) for the training phase of the proposed system.

For components $I$ and $Q$, the signal from the $s_1$th and $s_2$th reflecting surfaces to the selected $l$th receive antenna is expressed as follows:
\begin{eqnarray}\label{faded_signal2}
 y_l & = & \sqrt{E_x}\!\!\left(\left[\sum_{s_1=1}^{N} h_{s_1, l}^\Re \ e^{j \phi_{s_1}^\Re} \right] x_\Re \right.\nonumber \\
&  \quad + & \quad  \left. j \left[  \sum_{s_2=N/2+1}^{N}  h_{s_2, l}^\Im \ e^{j \phi_{s_2}^\Im}\right] x_\Im \right)\!+\! n_l,    
\end{eqnarray}
where \( l \in \{1,2, \dots, N_S\} \), the Rayleigh channel coefficients with mean zero and variance one are \( h_{s_1,l}^\Re =\alpha_{s_1,l}^\Re e^{-j \theta_{s_1,l}^\Re}  \) and \( h_{s_2,l}^\Im = \alpha_{s_2,l}^\Im e^{-j \theta_{s_2,l}^\Im}   \), representing the channels from the \( s_1 \)th and \( s_2 \)th reflecting surfaces to the \( l \)th receive antenna, respectively. $\mathbf{h}_{l}^\Re \in \mathbb{C}^{N/2 \times 1}=\left[h_{1,l}^\Re, h_{2,l}^\Re \ldots,  \ldots, h_{N/2,l}^\Re\right]$ and  $\mathbf{h}_{l}^\Im \in \mathbb{C}^{N/2 \times 1}=\left[h_{{N/2+1},l}^\Im, h_{{N/2+2},l}^\Im \ldots,  \ldots, h_{N,l}^\Im\right]$.  Where $\mathbf{h}_{l}^\Re$ and $\mathbf{h}_{l}^\Im$ are the $l$th column vectors of the $\textbf{H}_S^\Re$ and $\textbf{H}_S^\Im$ channel matrices. $n_l \sim \mathcal{CN}(0, N_0)$ is the additive white Gaussian noise (AWGN) variable with zero mean and $N_0$ variance. Also, the adjustable phase values for the reflecting surfaces $s_1$ and $s_2$ are $\phi_{s_1}^\Re=\theta_{s_1,l_\Re}^\Re$ and $\phi_{s_2}^\Im=\theta_{s_2,l_\Im}^\Im$, respectively.

The proposed system uses a maximum likelihood (ML) detector to estimate the selected receive antenna indices $l_\Re$ and $l_\Im$ and the symbols $x_\Re$ and $x_\Im$. The ML detector of the proposed system is expressed as follows:
\begin{equation}
\begin{aligned}
\Big[\hat{l}_{\Re},\hat{l}_{\Im},\hat{x}_{\Re},\hat{x}_{\Im} \Big] &= \arg\min_{{l}_{\Re},{l}_{\Im},{x}_{\Re},{x}_{\Im}} \sum^{N_S}_{l=1} \Bigg| y_l \\ - \sqrt{E_x} \Bigg[ \sum^{N/2}_{s_1=1} h_{s_1, l}^\Re e^{-j \varphi_{s_1, \ell_\Re}^\Re} {x}_{\Re} 
& \quad \!\!\!\!\!+ \!\!\! \sum^{N}_{s_2=N/2+1} \!h_{s_2,l}^\Im e^{-j \varphi_{s_2, l_\Im}^\Im} {x}_{\Im} \Bigg] \Bigg|^2.
\end{aligned}
\end{equation}
then, the bit demapper block in Fig. \ref{system_model} with the estimated $\hat{l}_{\Re}$, $\hat{l}_{\Im}$, $\hat{x}_{\Re}$, and $\hat{x}_{\Im}$  parameters generates the estimated data bits $\hat{\textbf{d}}$.

\begin{figure}[t] 
\centering
\includegraphics[width=0.45\textwidth]{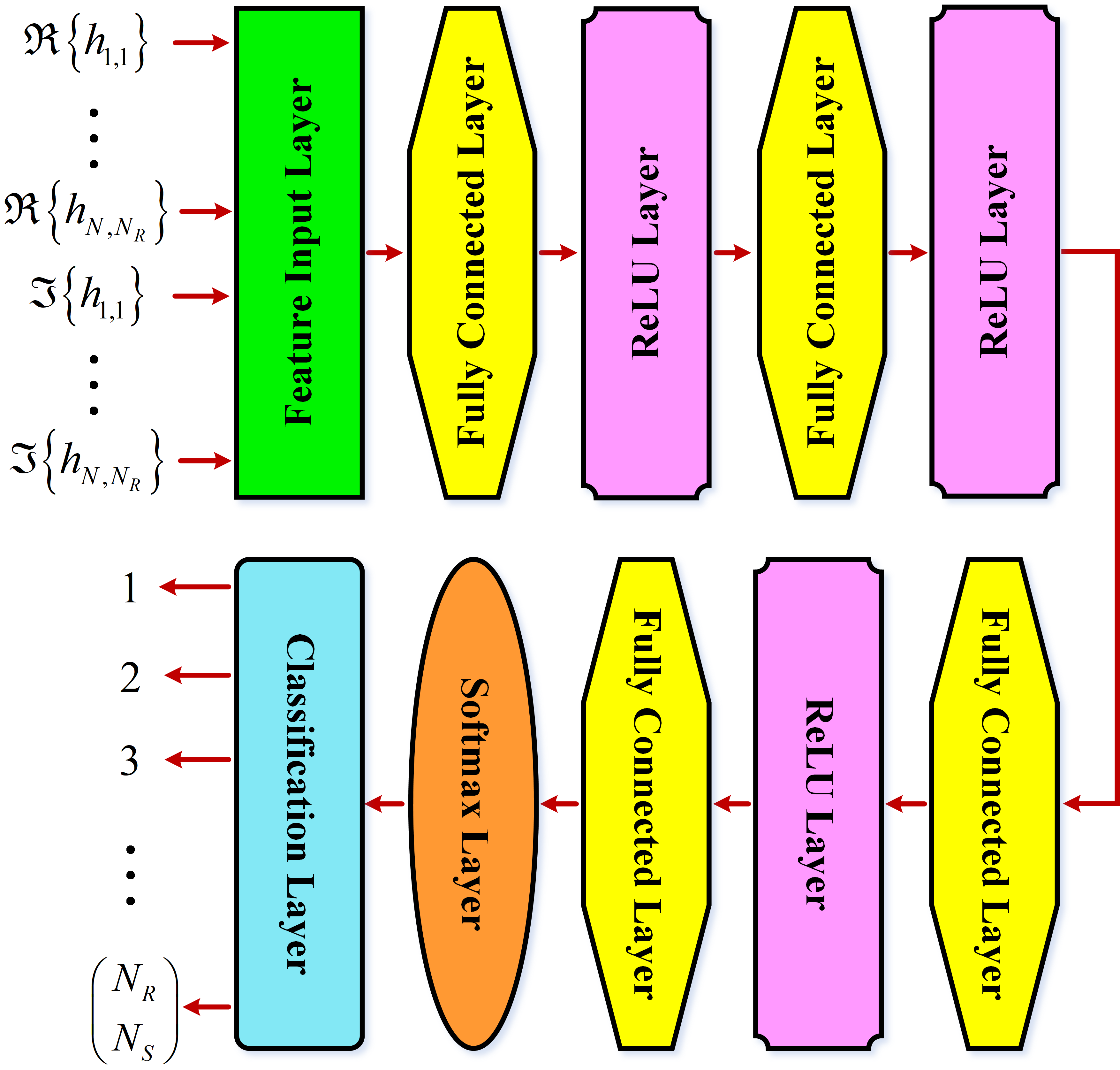}
	\caption{DNN architecture for the proposed system.}
	\label{mimari}
\end{figure}

The architecture of the DNN-based receive antenna selection algorithm is illustrated in Fig. \ref{mimari}\footnote{The input layer takes the real and imaginary channel components, and each of the four hidden layers is fully connected (256 neurons) with ReLU activation. The final softmax classification layer outputs one of the $\binom{N_R}{N_S}$ possible receive-antenna subsets corresponding to the best antenna selections determined by the COAS technique.}. In the proposed system, no preprocessing is performed, and the channel coefficient parameters are directly supplied to the DNN. In the proposed system, the feature vector for the COAS-based receive antenna selection is defined as:
\begin{equation}\label{input_v}
\mathcal{H} = \begin{bmatrix}
\Re\{\operatorname{vec}(\mathbf{H}_{\mathcal{S}})\} \\
\Im\{\operatorname{vec}(\mathbf{H}_{\mathcal{S}})\}
\end{bmatrix},
\end{equation}
where the vectorization operator is given by: $\operatorname{vec}(\mathbf{H}_{\mathcal{S}}) \triangleq \left[ h_{1,1},\, h_{2,1},\, \ldots,\, h_{N,1},\, h_{1,2},\, \ldots,\, h_{N,N_S} \right]^T$.

During the training stage, a feature vector is constructed, as defined in (\ref{input_v}), and random variables with the same distribution as the channel parameters are generated. The receiver may learn the true channel characteristics through channel estimation during training without prior knowledge of these parameters. Furthermore, labels for classification are produced as follows:
\begin{equation}
    \ell = \Bigg\{1, 2, \ldots, \binom{N_R}{N_S}\Bigg\}.
\end{equation}

Antenna selection is implemented by choosing \(N_S\) antennas from a total of \(N_R\) available receive antennas. For the specific case of \(N_R = 4\) and \(N_S = 2\), there are \(\binom{4}{2} = 6\) possible combinations. The labeling function \(\ell(\cdot)\) is defined as follows:
\begin{equation}
    \ell(\mathcal{S}) =
\begin{cases}
1, & \text{if } \mathcal{S} = \{1, 2\}, \\
2, & \text{if } \mathcal{S} = \{1, 3\}, \\
3, & \text{if } \mathcal{S} = \{1, 4\}, \\
4, & \text{if } \mathcal{S} = \{2, 3\}, \\
5, & \text{if } \mathcal{S} = \{2, 4\}, \\
6, & \text{if } \mathcal{S} = \{3, 4\}.
\end{cases}
\end{equation}
here, the set \(\mathcal{S}\) represents combinations of all possible antenna selections, where each subset contains \(N_S\) receive antennas selected from a total of \(N_R\) available receive antennas. As a result, for arbitrary \(N_R\) and \(N_S\), the mapping is defined over the set of all \(\binom{N_R}{N_S}\) possible antenna combinations, thereby assigning a unique label to each selected subset.

\begin{table}[h]
\centering
\caption{Training Parameters for the DNN-COAS-RIS-RQSM System}
\label{TP_DNN}
\begin{tabular}{ll}
\toprule
\textbf{Training Parameter} & \textbf{DNN} \\
\midrule
Number of Samples           & $10^6$  \\
Validation Split            & 10\%         \\
Mini-Batch Size                  & 256          \\
Number of Hidden Layers     & 4            \\
Iteration Steps             & 500          \\
Learning Rate               & 0.0005       \\
Optimization Algorithm      & Adam         \\
\bottomrule
\end{tabular}
\end{table}

It has been observed from the simulations that a dataset size of at least \(10^6\) samples is necessary to achieve satisfactory results. Given the large size of the dataset, using mini-batches is recommended to shorten the processing time by splitting the full data into smaller groups. In this study, a mini-batch size of 256 is determined to be appropriate. Furthermore, to ensure an efficient learning process, the learning rate is set to \(0.0005\), and the number of epochs is maintained at $400$. These settings are summarized in Table~\ref{TP_DNN}. Furthermore, the dataset used for the proposed system is generated according to the procedure outlined in Algorithm~1.

\definecolor{headerblue}{RGB}{100,100,100} 
\definecolor{lightgray}{RGB}{240,240,240} 

\begin{table*}[!ht]
\centering
\addtolength{\tabcolsep}{5pt}
\caption{Computational complexity comparisons of COAS and DNN techniques for the DNN-COAS-RIS-RQSM system.}
\label{tab_DNN_scenarios}
\rowcolors{2}{lightgray}{white} 

\begin{tabular}{@{}>{\bfseries}l c c c c c r r@{}}
\toprule
\rowcolor{headerblue} 
\color{white} \textbf{Cases} & 
\color{white} $\mathcal{L}$ & 
\color{white} $(S_1,\cdots,S_{\mathcal{L}})$ & 
\color{white} $N$ & 
\color{white} $N_R$ & 
\color{white} $N_S$ &
\color{white} COAS (\textbf{RMs}) &
\color{white} DNN (\textbf{RMs})
\\ 
\midrule
Case 1 & $2$ & $(4,\;4)$ & $8$ & $4$ & $2$ & $128$ & $296$ \\
Case 2 & $3$ & $(32,\;32,\;32)$ & $16$    & $4$ & $2$ & $256$       & $6336$       \\
Case 3 & $4$ & $(256,\;256,\;256,\;256)$ & $64$    & $8$ & $4$ & $2048$      & $476\,672$       \\
\bottomrule
\end{tabular}
\end{table*}

\SetAlgorithmName{Algorithm}{Alg.}{List of Algorithms}
\SetKwInput{KwIn}{Input}
\SetKwInput{KwOut}{Output}

\begin{algorithm}[h]
\SetAlgoLined
\caption{Data Generation for the Proposed DNN-Based System}
\KwIn{\(N,N_R, N_S\)} 

\For{\(i = 1 \to \text{dataset size}\)}
{
    \textbf{Step 1:} Generate a random \(N \times N_R\) sized channel matrix \(\mathbf{H}\) under the desired fading assumptions (e.g., Rayleigh).\\
    \textbf{Step 2:} Apply the COAS technique to select \(N_S\) antennas from the total \(N_R\) antennas, and assign a unique label for each combination:
    \[
      \ell = \Bigl\{1,2,\ldots,\binom{N_R}{N_S}\Bigr\}.
    \]
    \textbf{Step 3:} Record the channel matrix by concatenating its real and imaginary parts:
    \[
      \mathcal{H} = 
      \bigl[\,
        \Re\{\text{vec}(\mathbf{H})\}^T \,\,
        \Im\{\text{vec}(\mathbf{H})\}^T
      \bigr]^T.
    \]
}

Divide the generated dataset such that 10\% is set aside for validation purposes.

\KwOut{
\(\mathcal{H}^{(\text{train})}\): Training data, 
\(\mathcal{H}^{(\text{test})}\): Test data,
\(\ell^{(\text{train})}\): Training labels,
\( \ell ^{(\text{test})}\): Test labels
}

\end{algorithm}

\section{Complexity Analyses}
This section presents the DNN and COAS complexity for the proposed DNN-COAS-RIS-RQSM system.

\subsection{DNN Complexity Analysis}

The computational complexity of the proposed DNN-based scheme is assessed by considering the operations performed in each of its layers. In this scheme, each neuron applies a transformation and an activation function that linearly combines its inputs using weight and bias values. In the proposed system, the channel matrix $\mathbf{H}$ is of size $N \times N_R$. Since the antenna selection process is performed for this $N_R$ column, the entire matrix $\mathbf{H}$ is given as input to the DNN for optimal subset selection. Moreover, since the complex-valued $\mathbf{H}$ matrix is split into real and imaginary parts before being fed to the DNN, the number of input features is doubled. Therefore, the input dimension is as follows:
\begin{eqnarray}
\mathcal{P} \;=\; 2 (N N_R). 
\end{eqnarray}
Consider a neural network consisting of $\mathcal{L}$ hidden layers, $\mathcal{P}$ input neurons, and $\mathcal{T}=\binom{N_R}{N_S}$ output neurons. Let the $m^{\text{th}}$ hidden layer contain $S_m$ neurons. The total number of multiplication operations required throughout the network can be expressed as follows:
\begin{eqnarray}
\mathcal{O}_{\text{DNN}}=\left( \mathcal{P} S_1 + \mathcal{T} S_{\mathcal{L}} + \sum_{m=1}^{\mathcal{L}-1} S_m S_{m+1} \right).
\end{eqnarray}

\begin{figure}[t]
    \centering
    \includegraphics[width=1\linewidth]{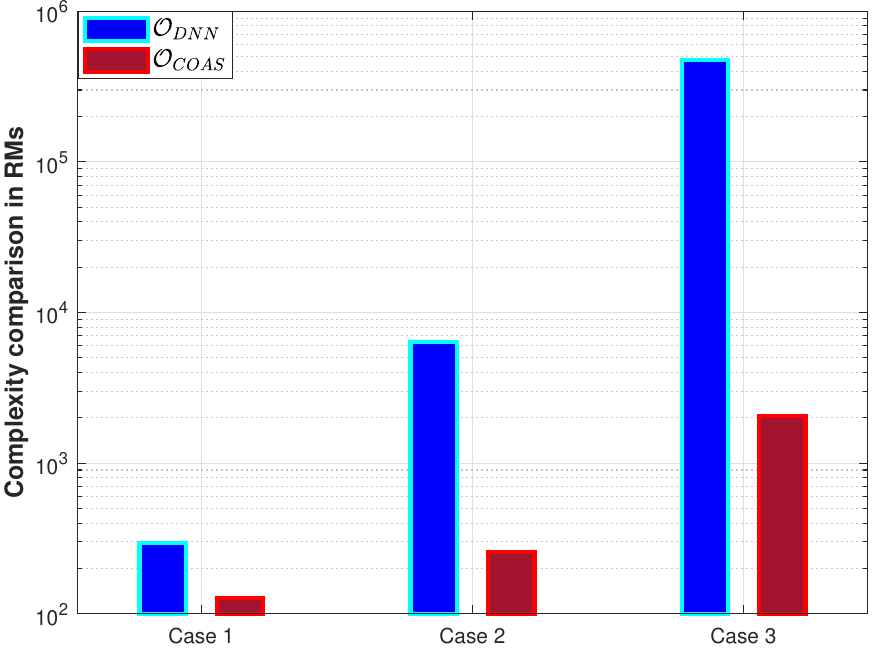}
    \caption{Complexity comparisons of COAS and DNN schemes for the DNN-COAS-RIS-RQSM system.}
    \label{fig_complexity_comparison}
\end{figure}

\subsection{COAS Complexity Analysis}
This section presents the COAS computational complexity analysis of the proposed system in terms of real multiplications (RMs). The Frobenius norm operation $||x_1 + jx_2||^2$ involves $4$ RMs, where $x_1$ and $x_2$ are any two real numbers. Also, the multiplication of two complex numbers requires $4$ RMs. In the COAS technique, the Frobenius norm operator in (\ref{coaseq1}) results in $4N$ RMs, and this operation is repeated $N_R$ times. Therefore, the computational complexity of the COAS technique is obtained as follows: 
\begin{eqnarray}
\mathcal{O}_{\text{COAS}} = 4NN_R \ \  \text{RMs}.
\end{eqnarray}

Table \ref{tab_DNN_scenarios} presents the computational complexity comparisons for the DNN and COAS techniques in the RIS-RQSM system. Three different cases are included in Table \ref{tab_DNN_scenarios}, each defined by specific system parameters. The same cases are also illustrated in Fig. \ref{fig_complexity_comparison}, where the computational complexities of both techniques are shown visually. Both Table \ref{tab_DNN_scenarios} and Fig. \ref{fig_complexity_comparison} show the higher computational complexity of the DNN technique compared to the COAS technique. Although the DNN technique involves higher computational complexity, it achieves improved error performance compared to the COAS technique. This indicates a trade-off between computational complexity and detection accuracy in the proposed system design.

\section{Simulation Results}
This section presents simulation results over Rayleigh fading channels, comparing the performance of the traditional COAS-RIS-RQSM approach with the proposed DNN-COAS-RIS-RQSM technique. In the simulations, the DNN is trained on $10^6$ samples, enabling reliable antenna selection and significantly reducing BER at moderate to high SNR levels. Also, the SNR in the simulation curves is expressed in decibels (dBs) and is denoted as $\mathrm{SNR\ (dB)} = 10\log_{10} \left( \frac{E_s}{N_0} \right)$, where $E_s$ represents the symbol energy.

\begin{figure}[t]
\centering
\includegraphics[width=0.47\textwidth]{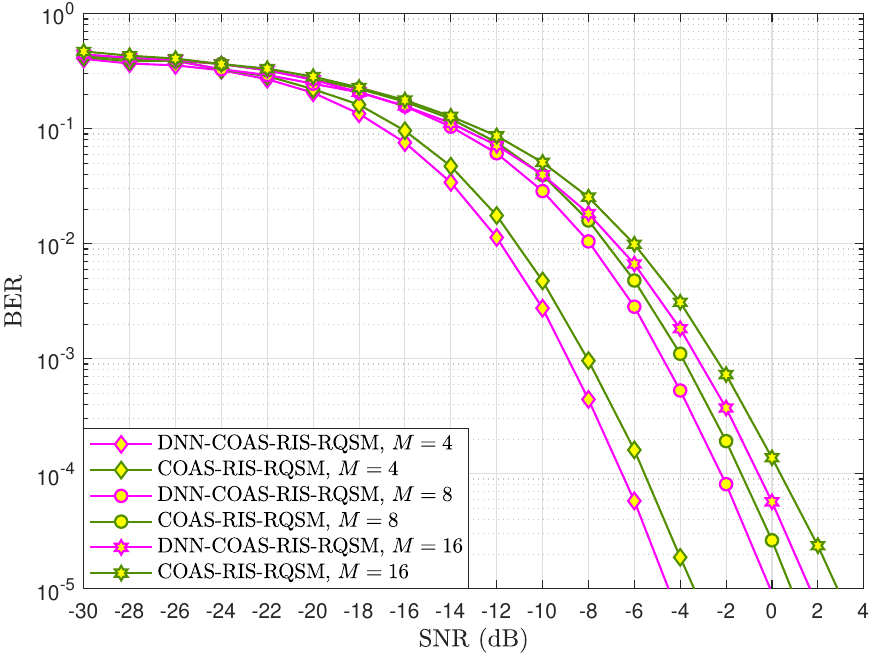}
\caption{BER performance comparison of DNN-COAS-RIS-RQSM and COAS-RIS-RQSM for \(N_R=4\), \(N_S=2\), \(N=16\), and \(10^6\) training samples, with different modulation orders \(M\).}
\label{MS}
\end{figure}

In Fig.~\ref{MS}, the BER performance of the proposed DNN-COAS-RIS-RQSM and COAS-RIS-RQSM is compared under the conditions of $N_R = 4$, $N_S = 2$, and $N = 16$ reflecting elements for three different modulation orders ($M = 4,\ 8,\ 16$). Fig.~\ref{MS} shows that the proposed DNN-COAS-RIS-RQSM system achieves better error performance than the conventional COAS-RIS-RQSM system. At the target BER of $10^{-5}$, an SNR gain of $1.12$ dB is achieved by the proposed DNN-COAS-RIS-RQSM scheme over the conventional COAS-RIS-RQSM approach when $M = 4$, while SNR gains of $0.77$ dB and $1.20$ dB are observed for $M = 8$ and $M = 16$, respectively.

\begin{figure}[t]
\centering
\includegraphics[width=0.47\textwidth]{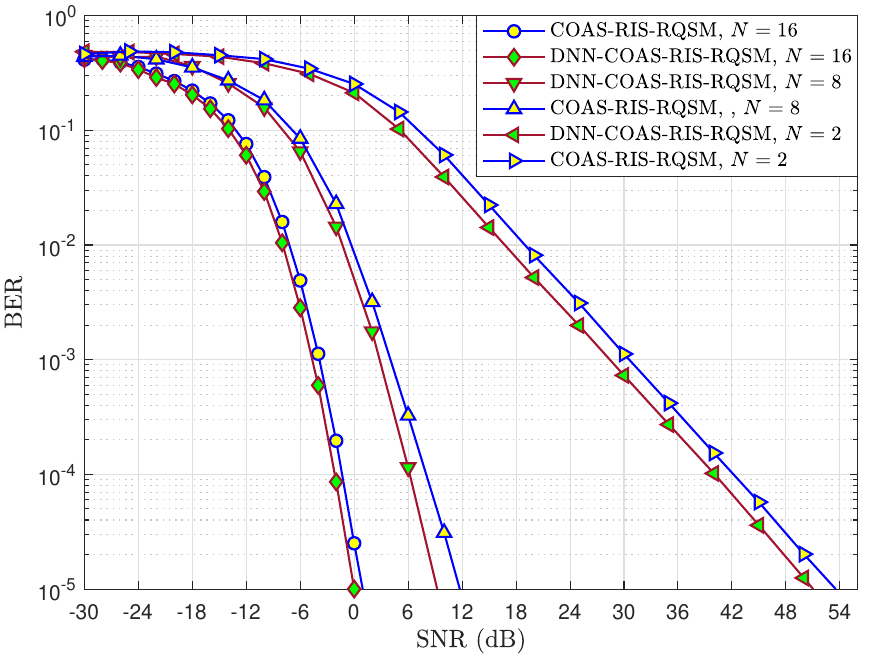}
\caption{BER performance comparison of DNN-COAS-RIS-RQSM and COAS-RIS-RQSM for \(N_R=4\), \(N_S=2\), \(M=8\), and \(10^6\) training samples, with varying numbers of reflecting elements \(N\).}
\label{NS}
\end{figure}

In Fig.~\ref{NS}, the BER performance of the proposed DNN-COAS-RIS-RQSM and COAS-RIS-RQSM systems is compared for $N = 16$, $N = 8$, and $N = 2$, while $N_{R} = 4$, $N_{S} = 2$, and $M = 8$.  Also, at a target BER of \(10^{-5}\), the DNN-COAS-RIS-RQSM system outperforms the COAS-RIS-RQSM system by an SNR gain of \(0.95\,\mathrm{dB}\) when \(N = 16\). This gain increases to \(2.52\,\mathrm{dB}\) for \(N = 8\), and remains at \(2.52\,\mathrm{dB}\) when \(N = 2\).


Overall, these simulation results confirm the effectiveness of integrating RQSM, COAS, and RIS under a deep-learning framework. By selecting the most advantageous subset of receive antennas and tuning the RIS phase shifts, the proposed DNN-COAS-RIS-RQSM scheme achieves consistently lower BER than its traditional counterpart for various system parameters. These results underscore the strong synergy between IM, DNN, and RIS, positioning the proposed system as a strong candidate for future wireless networks that demand high spectral efficiency and reliability.



\section{Conclusion}
This paper proposes a COAS-assisted DNN-based antenna selection method to enhance the RIS-RQSM system performance. Simulation results on Rayleigh fading channels demonstrate significant BER performance improvements of the proposed DNN-COAS-RIS-RQSM approach compared to the conventional COAS-RIS-RQSM method, particularly at medium to high SNRs. Furthermore, a comparative analysis of computational complexity between the DNN and conventional COAS techniques is also conducted, revealing a trade-off between complexity and detection accuracy. The proposed DNN-COAS-RIS-RQSM system offers low energy consumption, high spectral efficiency, and reliable low-error data transmission, outperforming traditional wireless communication methods and meeting the requirements of sixth-generation (6G) use cases such as reliable low-latency communication, IoT, and augmented reality.

\bibliographystyle{IEEEtran}
\bibliography{Referanslar.bib}

\end{document}